\documentclass[twocolumn]{aastex631}

\usepackage[version=4]{mhchem}
\usepackage{hyperref}

\newcommand \target{WISE~1828}
\newcommand \kzz{\ensuremath{K_{zz}}}

\newcommand \coiso{$^{13}\ce{CO}$}
\newcommand \coisoratio{$^{12}\ce{CO}/^{13}\ce{CO}$}
\newcommand \teff{$T_{\textrm eff}$}
\newcommand \mjup{$\textrm M_{J}$}
\turnoffedit
\turnoffeditone

\begin{document}
\title{High-precision atmospheric characterization of a Y dwarf with JWST NIRSpec G395H spectroscopy: isotopologue, C/O ratio, metallicity, and the abundances of six molecular species}

\submitjournal{The Astronomical Journal}
\accepted{Feb 7th, 2024}

\author[0000-0003-1487-6452]{Ben W.P. Lew}
\affiliation{Bay Area Environmental Research Institute, Moffett Field, CA 94035, USA}
\affiliation{NASA Ames Research Center, Moffett Field, CA 94035, USA}

\author[0000-0002-6730-5410]{Thomas Roellig}
\affiliation{NASA Ames Research Center, Moffett Field, CA 94035, USA}

\author[0000-0003-1240-6844]{Natasha E. Batalha}
\affiliation{NASA Ames Research Center, Moffett Field, CA 94035, USA}

\author{Michael Line}
\affiliation{School of Earth and Space Exploration, Arizona State University, Tempe AZ 85281, USA}

\author[0000-0002-8963-8056]{Thomas Greene}
\affiliation{NASA Ames Research Center, Moffett Field, CA 94035, USA}

\author{Sagnick Murkherjee}
\affiliation{University of California Santa Cruz, 1156 High Street. Santa Cruz, CA 95064, USA}

\author{Richard Freedman}
\affiliation{NASA Ames Research Center, Moffett Field, CA 94035, USA}
\affiliation{SETI Institute, 339 N Bernardo Ave Suite 200, Mountain View, CA 94043, USA}

\author[0000-0003-1227-3084]{Michael Meyer}
\affiliation{Department of Astronomy, University of Michigan, Ann Arbor, MI 48109, USA}

\author[0000-0002-5627-5471]{Charles Beichman}
\affiliation{NASA Exoplanet Science Institute, Infrared Processing and Analysis Center (IPAC)}
\affiliation{Jet Propulsion Laboratory, California Institute of Technology, Pasadena, CA 91125, USA}

\author{Catarina Alves De Oliveira}
\affiliation{European Space Agency, European Space Astronomy Centre, Camino Bajo del Castillo s/n, 28692 Villanueva de la Ca/~nada, Madrid, Spain}

\author[0000-0003-1863-4960]{Matthew De Furio}
\affiliation{Department of Astronomy, University of Texas at Austin,2515 Speedway, Stop C1400 Austin, Texas 78712-1205, USA}

\author[0000-0002-6773-459X]{Doug Johnstone}
\affiliation{NRC Herzberg Astronomy and Astrophysics, 5071 West Saanich Rd, Victoria, BC, V9E 2E7, Canada}
\affiliation{Department of Physics and Astronomy, University of Victoria, Victoria, BC, V8P 5C2, Canada}

\author[0000-0002-7162-8036]{Alexandra Z. Greenbaum}
\affiliation{IPAC, Caltech, 1200 E. California Blvd., Pasadena, CA 91125, USA}

\author{Mark Marley}
\affiliation{Lunar and Planetary Laboratory, University of Arizona, Tucson, AZ 85721, USA}

\author[0000-0002-9843-4354]{Jonathan J. Fortney}
\affiliation{University of California Santa Cruz, 1156 High Street, Santa Cruz, CA 95064, USA}

\author[0000-0002-6395-4296]{Erick T. Young}
\affiliation{Universities Space Research Association, 425 3rd Street SW, Suite 950, Washington DC 20024, USA}

\author[0000-0002-0834-6140]{Jarron Leisenring}
\affiliation{Steward Observatory, University of Arizona, Tucson, AZ 85721, USA}

\author[0000-0003-4850-9589]{Martha Boyer}
\affiliation{Space Telescope Science Institute, 3700 San Martin Drive, Baltimore, MD 21218, USA}

\author{Klaus Hodapp}
\affiliation{University of Hawaii, Hilo, HI,96720, USA}

\author{Karl Misselt}
\affiliation{Steward Observatory, University of Arizona, Tucson, AZ 85721, USA}

\author[0000-0003-2434-5225]{John Stansberry}
\affiliation{Space Telescope Science Institute, 3700 San Martin Drive, Baltimore, MD 21218, USA}

\author[0000-0002-7893-6170]{Marcia Rieke}
\affiliation{Steward Observatory, University of Arizona, Tucson, AZ 85721, USA}



\begin{abstract}

The launch of the James Webb Space Telescope (JWST) marks a pivotal moment for precise atmospheric characterization of Y dwarfs, the coldest brown dwarf spectral type.
In this study, we leverage moderate spectral resolution observations (R $\sim$ 2700) with the G395H grating of the Near-Infrared Spectrograph (NIRSpec) onboard of JWST to characterize the nearby (9.9\,pc) Y dwarf WISEPA J182831.08+265037.8.
With the NIRSpec G395H 2.88-5.12\,µm spectrum, we measure the abundances of \ce{CO, CO2, CH4, H2S, NH3, and H2O}, which are the major carbon, nitrogen, oxygen, and sulfur bearing species in the atmosphere. 
Based on the retrieved volume mixing ratios with the atmospheric retrieval framework CHIMERA, we report that the C/O ratio is $0.45 \pm 0.01$, close to the solar C/O value of 0.458, and the metallicity to be +0.30 $\pm$ 0.02 dex.
Comparison between the retrieval results with the forward modeling results suggests that the model bias for C/O and metallicity could be as high as 0.03 and 0.97 dex respectively.
We also report a lower limit of the  \coisoratio\  ratio of $>40 $, being consistent with the nominal solar value of 90. 
Our results highlight the potential for JWST to measure the C/O ratios down to percent-level precision and characterize isotopologues of cold planetary atmospheres similar to \target.

\end{abstract}



\section{Introduction} \label{sec:intro}

Y dwarfs, the coldest brown dwarf spectral type with temperatures below 500K \citep{cushing2011, Kirkpatrick2011}, offer a unique opportunity for studying the rich atmospheric chemistry and physics in cold giant planets.
The combination of low temperature and  molecular opacity of \ce{CH4,H2O, and NH3} causes the Y-dwarf spectrum to be dim in the near-infrared (1-2.5\,µm) while bright in the mid-infrared (3-5\,µm), coinciding with the opacity window of water vapor and methane \citep{marley2009}.
Mid-infrared observations are thus critical for mapping the spectral energy distribution and characterizing Y-dwarf atmospheric properties.
The ease of observations compared to cold giant planets orbiting bright stars makes Y dwarfs ideal exoplanet analogs.
By studying Y dwarf atmospheres, we can bridge the knowledge gap regarding how giant planets' atmospheric properties evolve across different temperatures, from the Jupiter-like temperature ($\sim$150K) to the warm ($>500\,\textrm K$) temperature that is found among T-dwarfs and exoplanets like 51 Eri-b. 
Furthermore, comparative studies of Y-dwarf and other giant planet atmospheres will shed light on potential connections between atmospheric properties and the formation and evolution histories of cold giant planets.

To date, approximately 50 Y-dwarfs have been discovered, primarily  through the Wide-field Infrared Survey Explorer (WISE) and Spitzer photometric observations \citep[e.g.,][]{kirkpatrick2019}.
Studies by \citet{morley2012,leggett2013,leggett2017} reveal that the color-magnitude relation (e.g., H-W2 vs. $\textrm M_{W2}$) of Y-dwarfs deviate from that of the chemical equilibrium cloudless models.
To better explain the Y-dwarf observed colors and absolute magnitudes, numerous atmospheric models have been developed to explore the possible chemical and physical properties, including water and sulfide clouds \citep{morley2012,lacy2023}, non-equilibrium chemistry \citep{phillips2020,mukherjee2022,lacy2023}, non-adiabatic thermal structure \citep{leggett2021}, non-solar metallicity \citep{marley2021}, and a non-solar carbon-to-oxygen (C/O) ratio \citep{cushing2021}. 
Despite the significant advance in atmospheric modeling of Y dwarfs, the existing discrepancy between observation and models underscores that our existing understanding of Y-dwarf atmospheres remains, as of yet, incomplete.

Spectroscopic data of Y dwarfs \citep[e.g.,][]{cushing2011,tinney2012,kirkpatrick2013,leggett2014,leggett2023} are essential for characterizing their atmospheres and testing atmospheric models of cold giant planets. However, spectroscopic observations of Y dwarfs are challenging due to their faintness in the near-infrared, the high thermal background, and the effects of telluric absorption in ground-based mid-infrared observations \citep[e.g.,][]{miles2020}.
\added{Space-based infrared spectroscopy is therefore vital for acquiring high-precision infrared spectra of Y dwarfs.}
One of the largest collections of Y-dwarf near-infrared spectra was obtained using the HST/WFC3 instrument by \citet{schneider2015}. Based on the dataset, \citet{zalesky2019} retrieved the temperature-pressure profiles, atmospheric composition, metallicity, and C/O ratios of 22 late-T and Y dwarfs.
A recent JWST spectroscopic study of a Y dwarf by \citet[][]{beiler2023} identifies the $\nu_3$ ammonia absorption features at 3\,µm.
These studies demonstrate the invaluable insights into Y-dwarf atmospheres gained from the spectroscopic data.

In this study, we utilize  JWST mid-infrared 3-5\,µm moderate-resolution (R $\sim$ 2700)  spectroscopy to characterize the atmospheric properties of the Y dwarf WISEPA J182831.08+265037.8. With the unprecedented sensitivity and mid-infrared spectral resolution of JWST, we aim to answer the following questions:
\begin{itemize}
    \item What are the abundances of the detected gas species in the Y-dwarf atmosphere?
\item How does the retrieved Y-dwarf atmospheric thermal structure compare to the radiative-convective equilibrium atmospheric models?
\item  Are there any spectral features of trace molecules and isotopologues present in the Y-dwarf spectrum?
\end{itemize}
\subsection{WISEPA J182831.08+265037.8}
\citet{cushing2011} reported the discovery of WISEPA J182831.08+265037.8 (hereafter \target) and categorized it as an archetypal Y-type brown dwarf. 
\citet{kirkpatrick2019} measured the parallactic distance of \target\ to be 
$9.93 \pm 0.23$ pc.
\edit1{\target\ is both a photometric and spectroscopic outlier, with an unusually red J-W2 colors compared to other Y dwarfs \citep[e.g.,][]{beichman2013,leggett2017}.}
The absolute magnitudes of \target\ are more luminous than the other Y dwarfs in both H and WISE W2 bands \citep[e.g.,][]{leggett2013,beichman2013}.

 Atmospheric modeling studies \citep[e.g.,][]{beichman2013,leggett2013,cushing2021} find that it is challenging to simultaneously fit models to the near-infrared YJHK and the mid-infrared (e.g., WISE W1 and W2) broadband photometry.
Given the unusual colors and overluminosity, \citet{beichman2013,leggett2017} suggested \target\ could be a tight binary system.
\citet{cushing2021} present the Hubble Space Telescope/Wide Field Camera 3 (WFC3) 1.1-1.7\,µm spectrum of WISE 1828.
They found that an atmospheric model with non-solar metallicity and C/O ratio provides a better fit to the near-infrared spectrum and mid-infrared photometry.
\citet{leggett2021} show that a binary system with a modified temperature-pressure atmospheric profile provides a decent fit to the observed photometry and spectrum.
Recently, \citet{defurio2023} reported no evidence of binarity with a separation beyond 0.5 au based on JWST NIRCam photometric observations.
\edit1{Consequently, the origins of the peculiar spectral and photometric features of \target\, remain elusive given the current data and models.}

\section{Observation and Data Reduction}
The JWST/NIRSpec observations of J1828 were executed on 2022 July 28, as part of the GTO Program PID 1189. 
At the beginning of the observations, a NIRSpec Wide Aperture Target Acquisition (WATA) image with a 3.6s exposure time and a clear filter was obtained to place the target at the slit center for fixed-slit spectroscopy.
The NIRSpec F290LP/G395H grating was used to obtain a J1828 spectrum ranging from 2.8 to 5.2\,µm under the NRSIRS2RAPID readout mode.
Three-point dithering along the slit S200A1 was performed during the observation. 
At each dither pointing, there were two integrations which were averaged over 26 groups each. 
The total exposure time for the G395H observation amounted to 2363.4 seconds.
For G395H spectra obtained with the S200A1 slit, there is a gap between detectors NRS1 and NRS2. 
The detector gap introduces a wavelength gap between 3.685 and 3.789\,µm in the 2.880-5.142\,µm spectra.

The data reduction was performed by mostly following the standard STScI pipeline with modified background subtraction, cosmic ray removal, and spectral extraction steps, described below.
The data were processed through the JWST pipeline version with a CRDS version of 11.16.21 under the context of \texttt{jwst\_1089.pmap}.
The data reduction steps are mainly done in three stages. 
In Stage 1, the \texttt{uncal.fits} files were processed through the superbias subtraction,  reference pixel correction, non-linearity correction, dark current correction, cosmic rays detection, ramp fitting step, and gain scale correction that eventually returned count-rate images with pixel values in the units of electrons per second. 
In Stage 2, the count rate images were assigned to the world coordinate system before proceeding to the customized background subtraction step.
For each exposure, a column-averaged background was estimated by taking the median of the pixel values at each wavelength in the cross-dispersion direction among the source-free regions. 
The variances of the median values over each column were treated as the uncertainty of the one-dimensional sky background and added to the Poisson noise per pixel.
After the background subtraction, the data was processed through wavelength correction, flat-fielding, and pathloss corrections, and was converted from an electron count rate to Janskys per steradian.
The dispersed spectra were then resampled and rectified so that the x-axis of the spectral images were equivalent to the spectral dispersion axis.
To identify any cosmic rays or bad pixels missed in the pipeline, we measured the centroid of each image column over a cross-dispersion aperture size of six pixels. A three-sigma clipping of the measured centroid values was used to flag the column numbers of spectral images that were possibly affected by cosmic rays or bad pixels.
Finally, an aperture size of six pixels was used in the spectral extraction step of the Stage 2 pipeline. The centers of the spectral traces for the three dithering positions were estimated as y= 9, 20, and 28 in the configuration file of spectral extraction step \texttt{extract\_1d}.

\subsection{Estimation of flat-field uncertainty}\label{sec:unc}
The wavelength-dependent flat-field uncertainty of NIRSpec was not yet available during our data reduction stage and is regularly being updated with more in-flight calibration data.
Based on the in-flight observation of a white dwarf, the NIRSpec G395H spectrum flux has an $0.91\pm 1.97\%$ RMS residual compared to the model flux template \citep{boker2023}.
The estimate is likely a lower limit because the reduction uses the flat-field reference file from the same dataset and does not include slitloss and systematic errors.
By comparing the spectra at different dithering positions, we found that the portion of spectra with estimated signal-to-noise ratios above 30 differ from each other systematically by around the 6\% level. 
The flat-field uncertainty at around 6\% is higher than the data uncertainty that is as low as 2\% based on the read noise and Poisson noise.
We therefore include an additional data uncertainty parameter as a nuisance parameter in our spectral fitting in Section \ref{sec:results}.

\section{Results} 
\begin{figure*}
    \centering
    \includegraphics[width=0.8\textwidth]{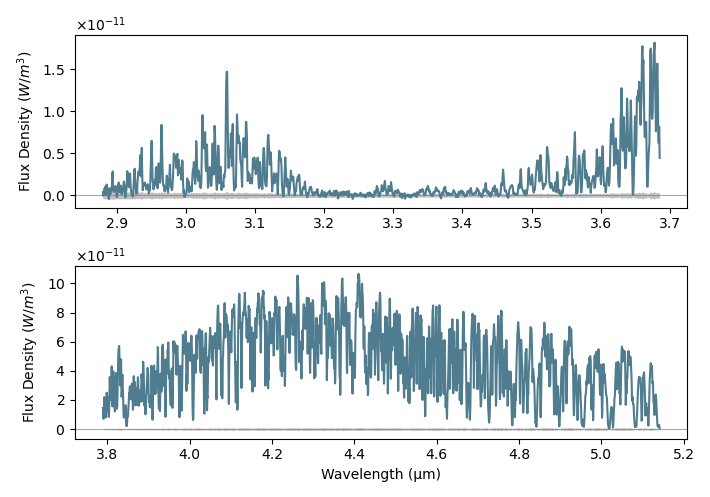}
    \caption{The reduced spectrum of WISE 1828. The parts of the spectra covered by the NRS1 and NRS2 detectors are shown in the top and bottom panels, respectively. The one-sigma uncertainties are plotted as grey shaded regions.}
    \label{fig:spectrum}
\end{figure*}

The median spectra averaged over three dithering positions are shown in Figure \ref{fig:spectrum}.
Based on the estimated noise that includes read noise, photon noise, and the background subtraction uncertainty, the peak of the spectra in the 4-4.5\,µm exceeds a signal-to-noise ratio (SNR) of 60  while the 3.00-3.10\,µm spectrum has SNRs of around 10. 
The spectrum shares overlapping wavelength coverage with the Spitzer [4.5] band. We integrate the G395H spectrum and derive a Spitzer [4.5]-band magnitude of $14.272 \pm 0.017$, which is 0.048 mag, or 1.8$\sigma$ higher than the measured Spitzer [4.5] mag of $14.32\pm 0.02$ \citep{kirkpatrick2019}. 
\edit1{Due to the data gap between 3.686-3.789 µm, we use the best-fit model spectra in Section 4.3 to interpolate the data to derive the equivalent [3.6] magnitude. We also assume that the interpolated flux shares a similar uncertainty to the mean data uncertainty near the gap (i.e. 3.636-3.686\,µm and 3.789-3.839\, µm). Based on these caveats, we derive a model-dependent [3.6] magnitude of $17.20 \pm 0.04$, which is about 0.285 mag dimmer than the \citet{kirkpatrick2019} value of $16.915\pm 0.05$. Future spectral observations with a complete [3.6] bandpass coverage will be helpful to provide a model-independent [3.6] magnitude measurement.}
\deleted{Therefore, the reduced JWST spectrum are consistent with the previous Spitzer [4.5] measurement.}

We then utilize two complementary atmospheric modeling tools to fit the G395H spectrum and to characterize the atmospheric thermal structure, compositions, C/O ratio, and metallicity in the following subsections.

\subsection{Description of atmospheric modeling tools} \label{sec:results}
Self-consistent radiative-convective equilibrium (RCE) models and atmospheric retrieval methods provide complementary insights into the atmospheric physics and chemistry of Y-dwarf atmospheres. 
In RCE models, the thermal and chemical structures are constrained by known physical processes. 
On the other hand, atmospheric retrievals are free to explore the parameter space which provides the best fit to the data. 
Therefore, cross-checking the retrieval results against the best-fit RCE models is useful for examining the possible physical and chemical processes and insuring that the retrieval results are not non-physical.
We leverage both the RCE models (Sonora Elf-Owl Grid, Murkherjee et al. in press) and an atmospheric retrieval framework (CHIMERA,\citealt{line2014a,line2014b}) to fit the spectrum and characterize the temperature, gravity, C/O ratio, metallicity, and atmospheric composition.

\subsubsection{Sonora Elf Owl models}
 The Sonora Elf Owl model grid (Murkherjee et al. in press) is a RCE cloudless atmospheric model grid that includes a self-consistent treatment of non-equilibrium chemistry of \ce{NH3,CO,H2O,CH4,CO2,N2, PH3} though 1D vertical mixing as described in \citet{mukherjee2022}.
 The model grid was computed with the \texttt{PICASO} atmospheric model \citep{batalha2019,mukherjee2023} and spans five parameters including effective temperature (\teff = [275,2400] K), gravity (log g $\mathrm  cms^{-2}$ = [3.25,5.5]), atmospheric metallicity ([M/H] =[-1,+1]), carbon-to-oxygen ratio (C/O = [0.22, 1.14]), and vertical eddy diffusion coefficient (log(\kzz (cgs))= [2,9]). The Elf Owl model grid assumes a constant {\kzz} throughout the atmosphere.  The high resolution spectra from the atmospheric model grid was computed using the resampled opacity grid at a wavelength resolution of 60,000 from \citet{picaso_resampled_highres}. For this work, the relevant line lists that were used to compute the opacities are CO \citep{li15rovibrational}, \ce{H2O} \citep{Polyansky2018H2O}, \ce{NH3} \citep{yurchenko11vibrationally,Wilzewski16}, \ce{CH4} \citep{Hargreaves2020ApJS}, \ce{CO2} \citep{HUANG2014reliable}. 
 There is one subtlety to the Elf Owl grid in how the \ce{PH3}  abundance is computed. The Elf Owl grid computed the \ce{PH3} abundance assuming quenching of the species for all but \ce{PH3}. For \ce{PH3} using the quenching methodology increases the abundance beyond chemical equilibrium. Because there is no observational evidence of strong \ce{PH3} absorption, the\ce{PH3} abundance is kept at chemical equilibrium. 
 We linearly interpolated the model spectra for parameter points in between the Elf Owl model grid points.
We then performed instrumental broadening, rotational broadening $v \sin i$, and wavelength shifting, or equivalently radial velocity shifting, on the model spectrum.
The instrumental broadening kernel is a wavelength-dependent function that is the same as the NIRSpec G395H spectral dispersion profile (\footnote{\url{https://jwst-docs.stsci.edu/jwst-near-infrared-spectrograph/nirspec-instrumentation/nirspec-dispersers-and-filters}}) and was scaled to match the unresolved emission line width of the calibration dataset (PID 1125). The rotational broadening was done using PyAstronomy \citep{pya}.
Finally, we scaled the model spectrum with a scaling factor of r$^2$/d$^2$, where r is the radius of the brown dwarf and d is the distance of 9.92 pc \citep{kirkpatrick2019}.
We use the ``MLFriends'' Nested Sampling Algorithm \citep{MLFriends2016, MLFriends2019} implemented in the open source code \texttt{UltraNest} \citep{Ultranest} and find best-fit solution by maximizing the log-likelihood function below:

\begin{equation}
    s^2 = \epsilon_{\mathrm{data}}^2 + (10^{\log(f)} \times \mathrm{data})^2
\end{equation}

\begin{equation}
    L = -0.5\sum \left[\frac{(\mathrm{data} - \mathrm{model})^2}{s^2} + 2 \pi s^2\right]
\end{equation}

where the $\epsilon$ is the flux uncertainty and $\log(f)$ is the logarithmic scaling factor for flux uncertainty. 
In total the \texttt{PICASO}/Elf Owl model fit has nine free parameters (five from the grid, and then the scaling factors for the flux uncertainty, radius, wavelength shift, and $v \sin i$). 

\begin{figure*}
    \centering
    \includegraphics[width=1.0\textwidth]{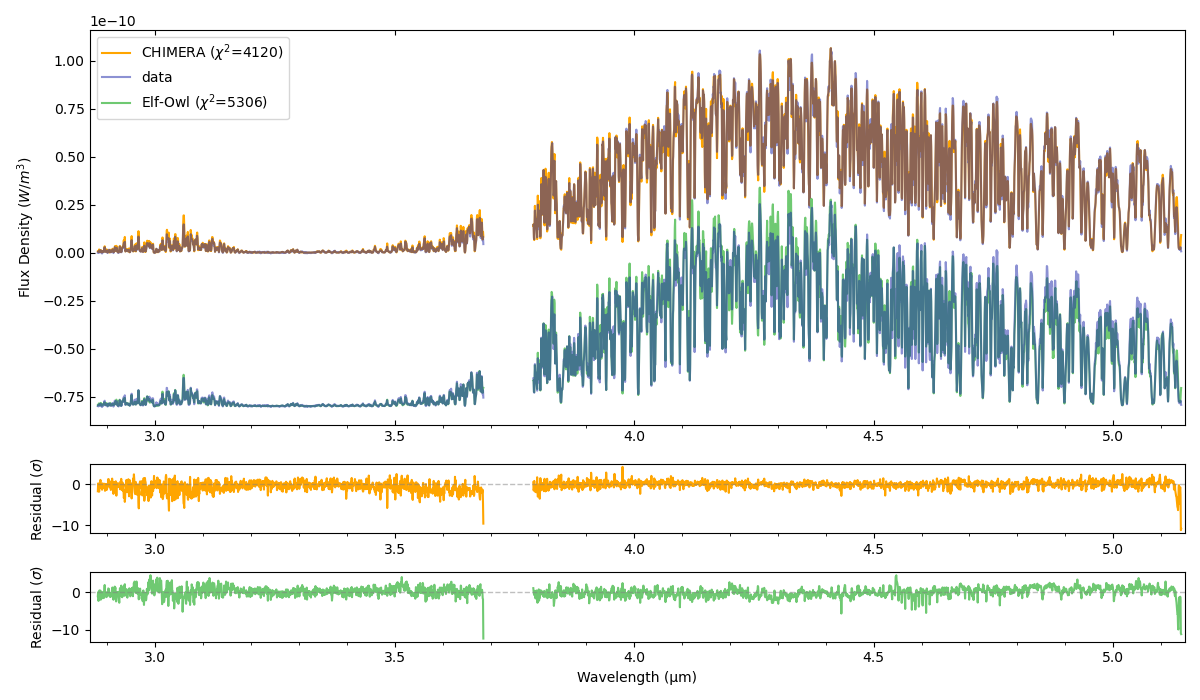}
    \caption{Top panel: the best-fit Elf-Owl model (semi-transparent green lines) and the CHIMERA atmospheric retrieval (semi-transparent orange lines) spectra. For clarity, the Elf-Owl model spectra and a copy of data spectra are shifted by $-8\times 10^{-11} \textrm Wm^{-3}$ in the y-axis. The original and shifted data spectra are plotted as semi-transparent blue lines for comparison with the CHIMERA and Elf-Owl models respectively. Middle and bottom panels: The residual for the CHIMERA and Elf-Owl models plotted in orange and green lines respectively. The sharp positive residual at 4.55\,µm of the Elf-Owl model is because of the lower-than-solar \ce{CH3D} abundance (See Section \ref{sec:trace}).}
    \label{fig:retrievalspec}
\end{figure*}

\subsubsection{CHIMERA}
CHIMERA \citep{line2013,line2014a,line2014b} is an atmospheric retrieval framework that has been tested on brown dwarf and exoplanet spectra \citep[e.g.,][]{line2015,zalesky2019,hood2023}. CHIMERA consists of 33 free parameters, including 15 pressure-temperature knots, gravity, radius, rotational broadening v $\sin i$, radial velocity, cloud vertical sedimentation efficiency ($f_{\textrm sed}$), cloud volume mixing ratio, cloud base pressure, and uncertainty inflation parameters, two hyperparameters for thermal structures ($\gamma$ and $\log(\beta)$), and constant volume mixing ratios of \ce{H2O, ^{12}CO, ^{13}CO, CO2, CH4, H2S, NH3}, and \ce{Na}. 
CHIMERA adopts the Markov chain Monte Carlo (MCMC) method to explore the parameter space and estimate the posterior probability distribution of the parameters. We refer to the details of the model setup to \citet{line2015}.

\subsection{Model Fitting Results}
The best-fit model spectra are plotted in Figure \ref{fig:retrievalspec}.
The best-fit CHIMERA and Elf-Owl model spectra give chi-squared values of 17180 and 42564 over 3233 data points respectively, corresponding to reduced chi-squares of 5.40 and 13.2.
We note that the apparent high reduced chi-squared is partially caused by the underestimated data uncertainty, which includes only photon, readout, and background subtraction noise, but not flat-field uncertainty.
Including an additional 6\% flat-field uncertainty (see Section \ref{sec:unc}) will lower the reduced chi-squared for the best-fit CHIMERA  and Elf-Owl models to 1.28 and 1.65, respectively.

In Figure \ref{fig:retrievalspec}, the NRS2 part of the spectra manifests larger residuals than in NRS1.
We observe that the residuals of the CHIMERA retrieval not only gives lower chi-squared but also appear to be less wavelength correlated than that of Elf-Owl model's residual.
This is not surprising because the CHIMERA model has over three times more parameters (33 vs. 10) than that of the Elf-Owl model.
The fitted cloud volume mixing ratio (VMR) by CHIMERA is low, with a log($VMR_{\textrm cloud}$) of $-12.23 \pm 5$. Therefore, we interpret that there is no significant cloud opacity contribution to the 3-5\,µm spectrum.
The fitted v $\sin i$ values of CHIMERA and Elf-Owl are lower than the resolution limit of NIRSpec/G395H ($\sim$110 km/s for R$\sim$2700) and should be interpreted as a free parameter that accounts for both instrumental and rotational broadening.

\edit1{Because of the unreasonably small formal uncertainties in the Elf-Owl model fitting results (see Table \ref{tab:relabundance}), we also use a bootstrapping method to estimate the uncertainties of the fitted parameters. We first resampled the residuals ($\sigma$) of the best-fit Elf-Owl models to generate 1000 model spectra with injected resampled noise. We then use the least chi-square method to obtain the best-fit parameters. Based on these 1000 fitting results, we obtain a 68-percentile interval as the 1$\sigma$ uncertainties which are listed in Table \ref{tab:relabundance}.}

In the following subsections, we focus on  the fitted thermal structure, atmospheric composition, the C/O ratio,  metallicity, and isotopologue constraints of the best-fit CHIMERA and Elf-Owl models.
We then inspect the residuals and search for potential minor species in Section \ref{sec:trace}.

\begin{figure*}
    \includegraphics[width=1.0\textwidth]{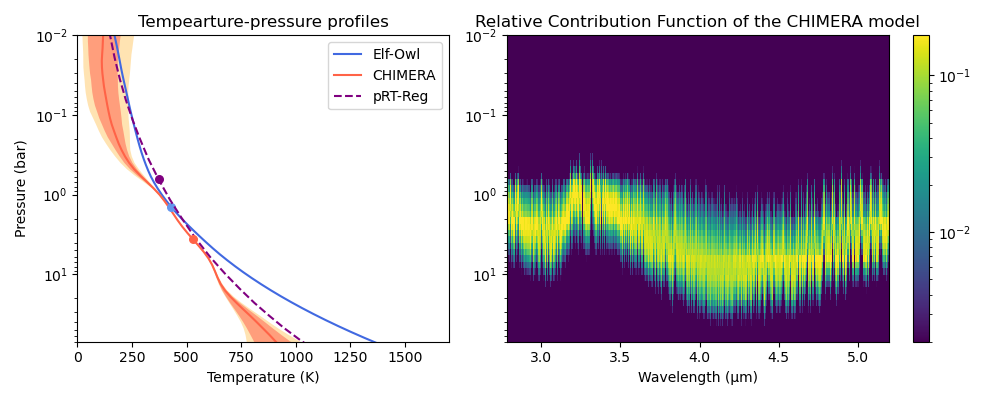}
    \caption{Left panel: the temperature-pressure profile of  CHIMERA  (\teff = $534^{+8}_{-25} $K, log(g) = $5.02 \pm 0.01$), the best-fit Elf-Owl model (\teff = $425 ^{+0.5}_{-0.3}$ \,K, log(g) = 4.38 $\pm$ 0.01), and of the petitRADTRANS retrieval  in \citet{barrado2023}. The colored dots indicate the pressures at which the local temperatures equal to the effective temperatures of the models. See Section \ref{sec:b23} for the discussion of the thermal structure.  Right panel: the contribution function of CHIMERA. The 3-5$\mu$m spectrum probes around 0.8-27 bar. The probed pressure range corresponds to the region in which the temperature-pressure profiles have small uncertainties.}
    \label{fig:contributionfunction}
\end{figure*}

\begin{figure}
    \includegraphics[width=0.48\textwidth]{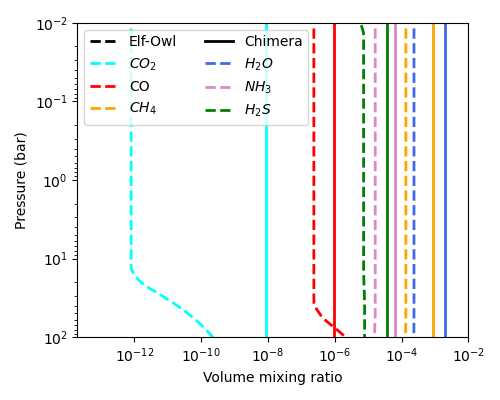}
    \caption{The fitted chemical abundances of the forward model Elf-Owl and the CHIMERA retrieval framework. The dashed and solid lines show the abundance of the best-fit model spectrum from the Elf-Owl models and CHIMERA,  respectively.}
    \label{fig:picaso_abundance}
\end{figure}

\subsection{Thermal structure and Contribution Function}

We plot the retrieved temperature-pressure (TP) profile in Figure \ref{fig:contributionfunction} with 68\% and 95\% confidence intervals.
As indicated in Figure \ref{fig:contributionfunction}, the TP profile is tightly constrained in the 1-20~bar region.
By integrating the spectra from 1 to 200\,µm based on the CHIMERA retrieval results, we estimate that the effective temperature is $534^{+8}_{-25} $\,K. 
We note that our best-fit model indicates that the G395H spectrum covers about 47\% of the bolometric luminosity.

Comparison of the TP profiles fitted with different modeling approaches provides a qualitative assessment on the potential systematic uncertainty due to the model assumptions.
In the left panel of Figure \ref{fig:contributionfunction}, we also plot the TP profile of the best-fit Elf-Owl model.
The best-fit Elf-Owl model has an effective temperature of $425^{+0.46}_{-0.33}$ \,K and a log(g) of $4.3 \pm 0.01$. 
The uncertainty of the Elf-Owl models are much smaller than the grid spacing and likely underestimated. 
Given a fixed pressure, the temperature of the CHIMERA retrieval is cooler than that of the Elf-Owl model.
Qualitatively, the two TP profiles are relatively similar in the 1-5 bar region and increasingly differ at higher pressures.
Therefore, caution should be exercised when interpreting the fitted TP profiles at a pressure of five bars or higher.

To illustrate the pressures from which the flux at different wavelengths is emitted, we use PICASO to calculate the contribution function based on the fitted atmospheric composition and thermal structure by CHIMERA and Elf-Owl models. In Figure \ref{fig:contributionfunction}, we show the relative contribution function of the CHIMERA retrieval, which is normalized by the flux density. 
Based on the calculated contribution function, the G395H spectrum probes 0.6-19 bar region based on the best-fit Elf-Owl model (not shown in Figure \ref{fig:contributionfunction}) and probes around the 0.8-27 bar region based on the CHIMERA retrieval results. 
Therefore, the fitted Elf-Owl and CHIMERA model spectra probe similar pressure ranges.
The contribution function results are also consistent with the tight constraints of the CHIMERA-retrieved TP profile in Figure \ref{fig:contributionfunction}.

\subsection{Atmospheric Composition}\label{sec:composition}
\begin{figure*}
    \centering
    
    \includegraphics[width=1\textwidth]{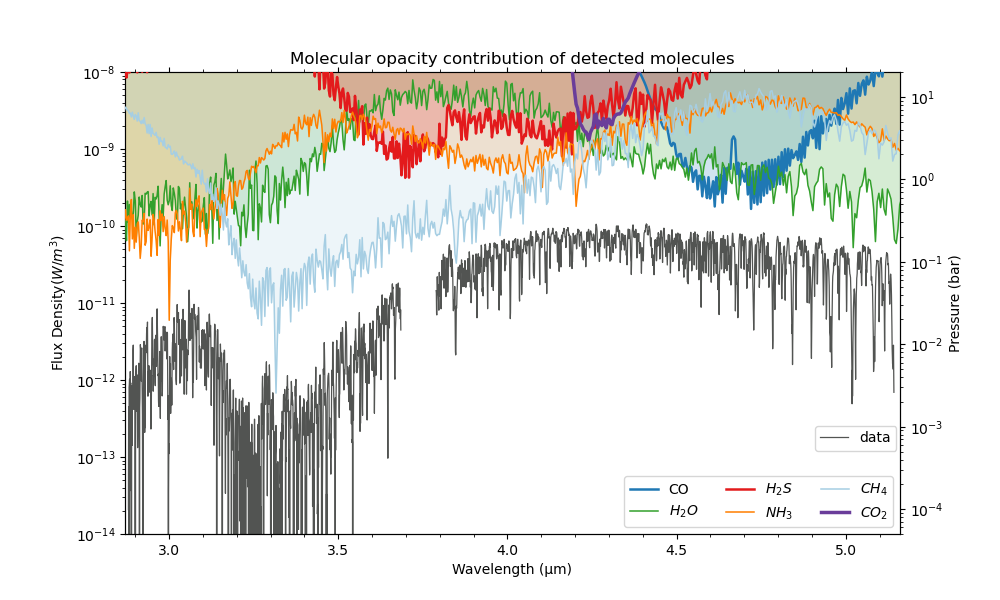}
    \caption{The spectrum is plotted in dark-grey lines along with the P($\tau$=0.01) of the detected molecules. The left y-axis shows the flux density of the spectrum while the right y-axis shows the photospheric pressure at which the molecular optical depth reach 0.01. The three most abundant molecules \ce{CH4, H2O, and NH3} are plotted in thin light blue, green, and orange lines respectively. The three less abundant molecules \ce{H2S,CO2,and CO} are plotted in thick red, purple, and blue lines respectively. The P($\tau$=0.01) lines of molecules are binned down into resolution of R=1350 for clarity.}
    \label{fig:zoomspectrum}    
\end{figure*}

\begin{table*}[!h]
\centering
\caption{\\The fitted key model parameters of Elf-Owl grid models and the CHIMERA retrieval framework }\label{tab:moleculelist}  
\begin{tabular}{cccc}
 \hline \hline
Model Parameters & CHIMERA & Elf-Owl & Bootstrapping 68 \% C.I. \\ \hline
$T_{\mathrm{eff}}$ (K) & $534^{+8}_{-25}$ & $425^{+0.46}_{-0.33}$ & [344, 436]  \\
Radius ($R_{\mathrm{Jup}}$) & 1.23 & $1.030 \pm 0.003$ & [0.82, 1.32] \\
    $\log(\textrm{g})$                                       &         $5.20 ^{+0.01}_{-0.02}$                   &     $4.38 \pm 0.01$ & [4.12, 4.75]                            \\
RV (km/s) &                             $-32.0\pm 0.15$         &    $-34.1\pm0.2$        & [-35, -34]   \\
v sin i (km/s)                               &   $60.20^{+0.19}_{-0.15}$                         &   $0.14^{+0.90}_{-0.11}$      & [0.017, 19]                     \\
$\log(K_{\textrm{zz}}$ (cgs))                               &   -                         &  $4.65 \pm 0.04$       & [4.55, 6.63]                     \\
log(CO)                                    &   -                         &   $-6.62 \pm 0.01 $                     \\

log($^{12}$CO)                                    &   $-6.01\pm 0.02$                         &   -                     \\
$\log (^{13}\textrm{CO}/^{12}\textrm{CO})$                                       &  $<-1.61$                           &    -                         \\
 (equiv. $^{12}$CO/$^{13}$CO) & $>40$ & - \\
log(\ce{CH4})                                          &    $-3.07^{+0.01}_{-0.02}$                        &  $-3.87 \pm 0.04$                              \\
log(\ce{CO2})                                          &    $-8.79^{+0.03}_{-0.04}$                        & $-12.09 \pm 0.01$                               \\
log(\ce{H2O})                                          & $-2.71^{+0.01}_{-0.02}  $                          & $-3.62 \pm 0.01$                              \\
log(\ce{NH3})                                          &   $-4.21 \pm 0.02$                         &   $-4.789 \pm 0.005$                           \\
log(\ce{H2S})                                          &   $-4.44 \pm 0.03$                         &    $-5.136 \pm 0.004$          \\
log(\ce{PH3}) & - & $-11.84 \pm 0.03$\\
\hline                    
\end{tabular}
\raggedright  

{\textbf{Note:} The molecular abundances of the Elf-Owl grid models are not free parameters and are altitude dependent. We listed the Elf-Owl model abundances at five bars, which is the averaged peak locations of the contribution function in Figure \ref{fig:contributionfunction}. The molecular abundances of CHIMERA are constant with altitude. \added{See the text in Section \ref{sec:results} for the description of the estimated 68\% confidence intervals (C.I.) for the Elf-Owl model parameters with a bootstrapping method.} }
\end{table*}

\begin{table}[]
\caption{The relative chemical abundances of CHIMERA and the Elf-Owl models }\label{tab:relabundance}
\begin{tabular}{ccc}
\hline \hline
                         & \multicolumn{1}{c}{CHIMERA}  & Elf-Owl (5 bar) \\ \hline
log(\ce{CH4/H2O}) & {$-0.36 \pm 0.02$ }    & $-0.25\pm 0.04$       \\
log(\ce{CO/H2O})  & {$-3.30^{+0.03}_{-0.02}$}     & $-3.00 \pm 0.01$    \\
log(\ce{CO2/H2O}) & {$-6.08 \pm 0.04$}    & $-8.47\pm0.01$              \\
log(\ce{NH3/H2O}) & {$-1.50^{+0.03}_{-0.02}$}    & $-1.17 \pm 0.01$\\
log(\ce{H2S/H2O}) & {$-1.73^{+0.04}_{-0.03}$}    & $-1.52 \pm 0.01$\\ \hline 
\end{tabular}
\end{table}

 In Figure \ref{fig:picaso_abundance}, we plot the best-fit molecular abundances of the CHIMERA and Elf-Owl models.
 We also plot the pressure at which the optical depth of individual molecule reaches 0.01 in Figure \ref{fig:zoomspectrum} to visualize the contribution from each molecule on the observed absorption feature.
 The \deleted{full} corner plot and marginalized distribution of the \deleted{33}\added{key} parameters are shown in Appendix \ref{sec:appendix}.
 In Table \ref{tab:moleculelist}, we list the median values and 1-$\sigma$ confidence intervals of the fitted molecular abundances.
 Guided by the marginalized distribution of molecules that show bounded lower and upper limits, we report the detection of \ce{H2O, CH4,^{12}CO,NH3,H2S, and CO2} in the G395H spectra.

\begin{figure*}
    \includegraphics[width=1.0\textwidth]{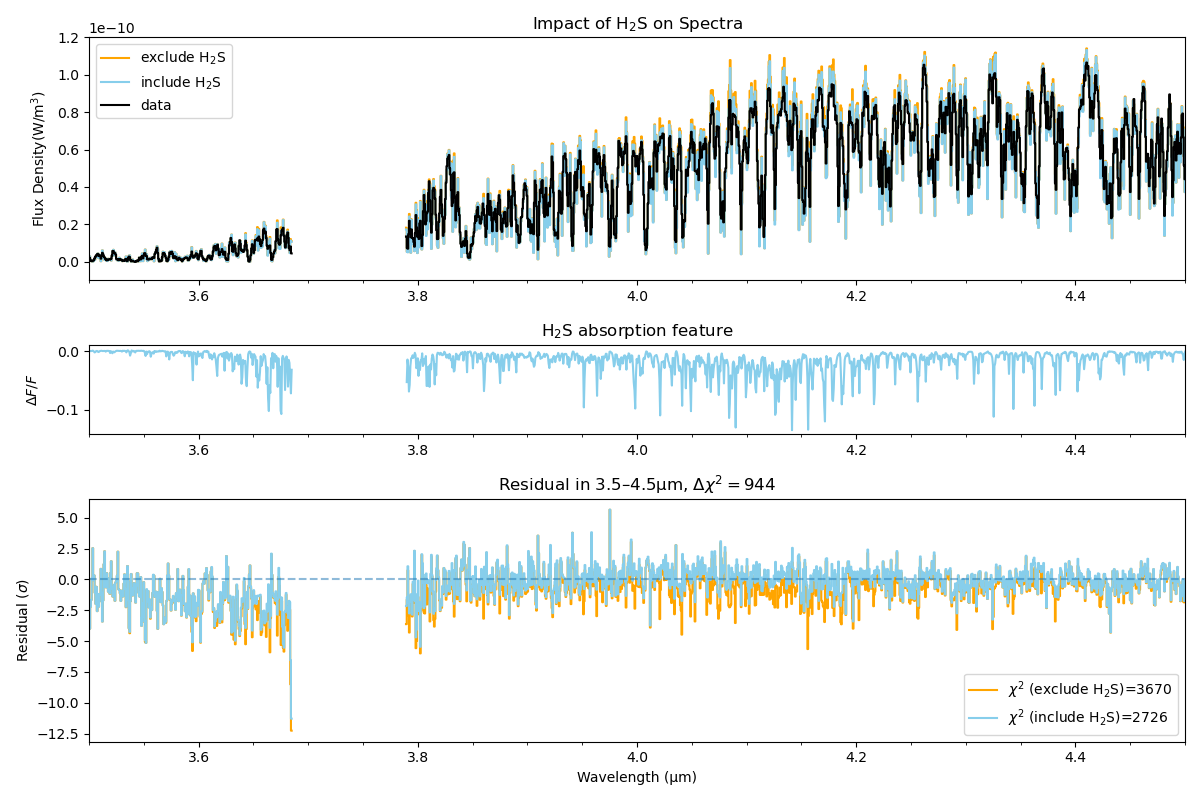}
    \caption{Top panel: The best-fit CHIMERA model spectrum with and without \ce{H2S} opacity are plotted in blue and orange lines, respectively.The data is plotted in black line. Middle panel: The fractional flux density change of the best-fit spectrum after the removal of \ce{H2S} opacity. Bottom panel: the residuals relative to the data uncertainties for the CHIMERA retrieval results  with and without \ce{H2S} opacity are plotted in blue and orange lines. Exclusion of \ce{H2S} opacity leads to an increase of $\chi^2$ by over 900.}\label{fig:h2s}
\end{figure*}

 To our knowledge, the only previous detections of \ce{H2S} in ultracool brown dwarfs are \citet{tannock2021,hood2023} with ground-based medium-to-high resolution near-infrared spectroscopy. 
 In Figure \ref{fig:h2s}, we plot the increase of the residuals with the removal the \ce{H2S} from the CHIMERA retrieval.
Our study presents one of the first bounded constraints on the \ce{H2S} abundance in a brown dwarf atmosphere.
On the other hand, the posterior distribution of $^{13}$CO/$^{12}$CO ratio of the CHIMERA retrieval does not show a clear lower bound, so we report no detection for \ce{^{13}CO} in the data.
For the best-fit Elf-Owl model that includes \ce{PH3}, we notice that the absorption by \ce{PH3} that is in chemical equilibrium causes the model to underestimate the flux density in the 4.1-4.3\,µm region. 
Meanwhile, the residuals of the CHIMERA retrieval, which does not include \ce{PH3}, does not exhibit spectral features similar to that expected from \ce{PH3} absorption at 4.3\,µm. 
Therefore, we report no evidence of \ce{PH3} absorption in the \target\ spectra.

Overall, the CHIMERA retrieval gives higher ( $>1 \sigma$) molecular abundances of \ce{H2O, CH4,^{12}CO,NH3,H2S, and CO2} when compared to the best-fit Elf-Owl model grid.
In particular, the \ce{CO2} abundance, which is sensitive to the metallicity \cite[e.g.,][]{zahnle2014}, of the best-fit Elf-Owl model is almost five orders of magnitude lower than that of the CHIMERA retrieval (VMR(\ce{CO2} of  $10^{-8.8}$ vs. $10^{-12}$).
Upon inspection of the residuals in the 4.2-4.3\,µm at which \ce{CO2} absorption occurs, we find that CHIMERA retrieval residuals are lower than the Elf-Owl model. 
Furthermore, the residuals of the Elf-Owl model in the 4.1-4.3\,µm is dominated by the excess \ce{PH3} absorption that overlaps with the \ce{CO2} absorption in the 4.2-4.3\,µm region.
Therefore, we argue that the fitted \ce{CO2} abundance by the CHIMERA retrieval is more reliable than that by the Elf-Owl model.

In Table \ref{tab:relabundance}, we derive the relative molecular abundances, which is likely to be less sensitive to the potential degeneracy between the gravity and molecular abundance.
Because the molecular abundance changes with altitude in the Elf-Owl models, we listed the molecular abundances at five bars, which is the averaged photospheric pressures probed by the 3-5\,µm spectrum.
The differences in the relative abundances are smaller than that in the absolute abundances between the two models.
Therefore, the relative molecular abundances are more robust against the model assumptions to reproduce the observed spectral features.

\subsubsection{Elemental Abundance}
We can derive atmospheric elemental abundances based on the fitted molecular abundances from the CHIMERA retrieval, while making certain assumptions about the background undetected gas species.
We do not do this for the fitted Elf-Owl models because the elemental abundances of forward models are set by the assumed metallicity and C/O ratio;
For the CHIMERA retrieval framework, we can derive the elemental abundance [C/H],[N/H], [O/H], and [S/H] based on the detected molecular abundances \ce{CH4,CO2,CO,NH3,H2S,and H2O}.
We calculate the hydrogen,carbon,nitrogen,oxygen abundance by the following equations:

\begin{equation}
\begin{aligned}
H &= \ce{2H2} + \ce{4CH4} + \ce{3NH3} + \ce{2H2O} + \ce{2H2S} \\
C &= \ce{CO2} + \ce{CO} + \ce{CH4} \\
S &= \ce{H2S} \\
N &= \ce{NH3} \\
O &= \ce{H2O} + \ce{CO} + \ce{2CO2} 
\end{aligned}
\end{equation}

We then normalize the abundances by the corresponding solar abundance (e.g., [C/H] = $\textrm[C/H]_{measured}$ - $\textrm[C/H]_{solar}$)
 The normalized elemental abundances are [C/H] of $+0.24^{+0.01}_{-0.02}$, [O/H] of $+0.34 ^{+0.01}_{-0.02}$, [N/H] of $-0.31 \pm 0.02$, and [S/H] of $+0.14 \pm 0.03$.

In addition to the detected molecules, there are other non-detected molecules that are also important as the potential reservoir of C,N,S,and O elements.
For example, the derived N/H is probably underestimated because of the missing \ce{N2} abundance. 
Based on the chemical abundance of the best-fit Elf-Owl models, the \ce{N2} abundance is about 60\% of \ce{NH3} in the 0.1-20 bar region.
Therefore, the contribution from \ce{N2} could increase the estimated [N/H] to +0.04, which is near solar [N/H] value.
We refer to Section \ref{sec:b23} for further discussion about ammonia abundances and nitrogen isotopologue.
The [O/H] is also likely underestimated because of silicate cloud (e.g., \ce{Mg2SiO4}) formation, which could lower the detected [O/H] by $\sim$30\% \added{and increase the C/O ratio by $\sim 40\%$ \citep[e.g.,][]{zalesky2019,calamari2022}. }
After considering the effect of other non-detected molecules, our results suggest that \target\ has above-solar [O/H], [S/H], and [C/H] abundances and likely near-solar [N/H] abundance.

\subsubsection{C/O ratio and metallicity}
\begin{figure*}
    \centering
    \includegraphics[width=0.9\textwidth]{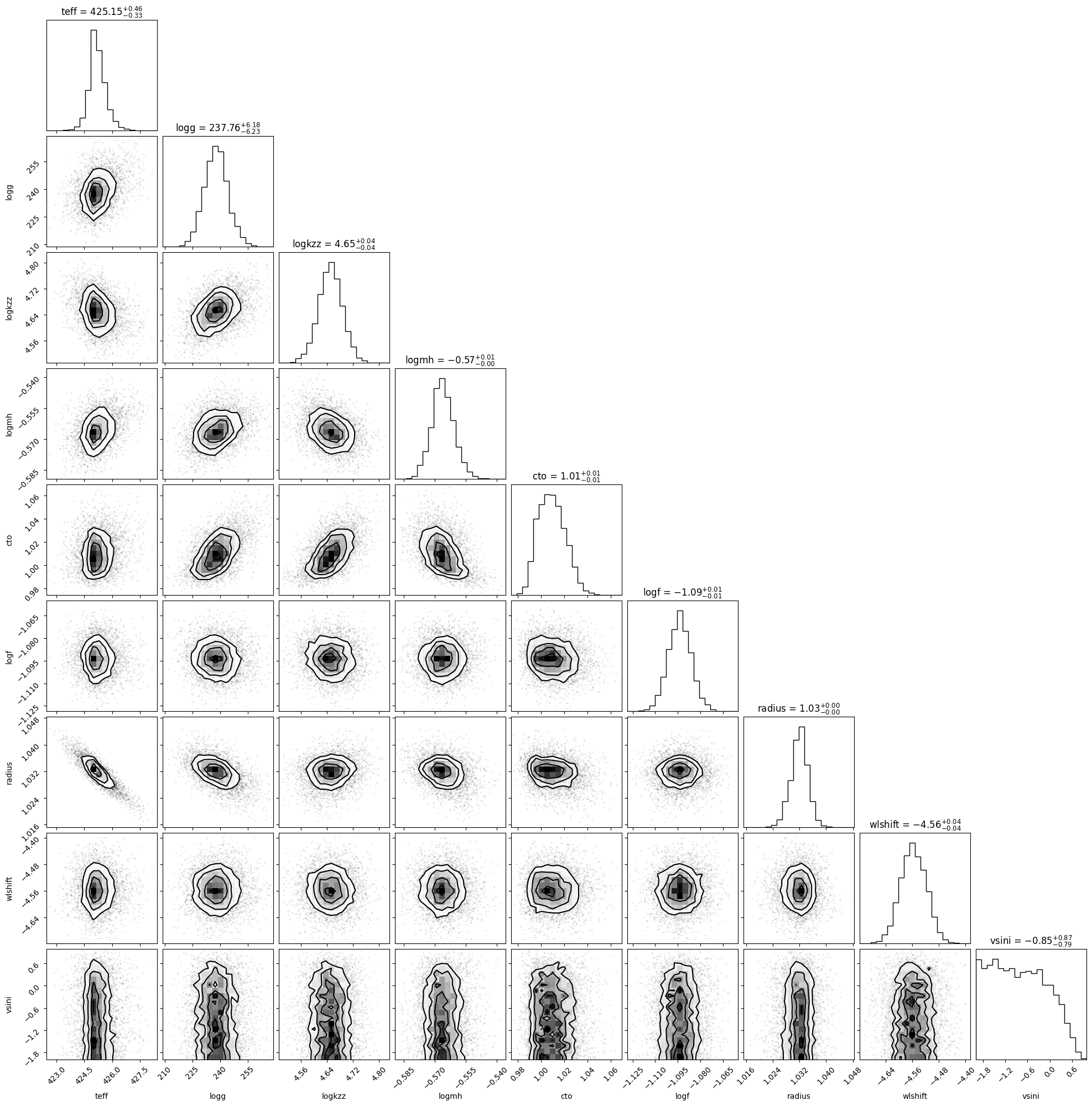}
    \caption{The corner plot of the Elf-Owl model grid fitting results. The nine model parameters are effective temperatures, gravity ($\log(g)$), vertical mixing $\log(K_{zz})$, metallicity log(M/H), C/O ratio, radius ($R_{\textrm Jup}$), wavelength shift in unit of $\AA$, and $\log$(v sini km/s).}
    \label{fig:elf-owl_corner}
\end{figure*}

The best-fit Elf-Owl model has a metallicity of [M/H]= $-0.57 \pm 0.01$ and a C/O ratio of 1.01$\times$Solar C/O value (0.458) \citep{Lodders2009}. The corner plot of the grid-model fitting shows that the metallicity is negatively correlated to the C/O ratio, as shown in Figure \ref{fig:elf-owl_corner}.
Based on the elemental abundances of the CHIMERA retrieval, we calculate the C/O ratio to be $0.43\pm 0.01$, similar to the solar C/O value. 
We also calculate the metallicity by summing the elemental abundance heavier than hydrogen. The derived metallicity [M/H] is equal to $0.30\pm 0.01$ dex, or about two times solar metallicity. 
Therefore, the two models share similar C/O ratios, but the Elf-Owl model has a metallicity 0.87 dex lower than the CHIMERA retrieval.

Similar to the estimation of oxygen abundance,  our estimation of C/O and metallicity did not account for the potential silicate condensate cloud formation. Therefore, the actual oxygen bulk abundance and the corresponding inferred metallicity should be higher than +0.30 dex while the C/O ratio should be lower than 0.43.
We note that the retrieved molecular abundances and metallicity by CHIMERA are correlated with gravity. We further discuss the interpretation and implications of the atmospheric composition, derived metallicity, and C/O ratio in Section \ref{sec:caveats} and \ref{sec:cto}.

\subsubsection{$^{13}$CO/$^{12}$CO ratio}
\begin{figure*}
    \centering
    \includegraphics[width=1.\textwidth]{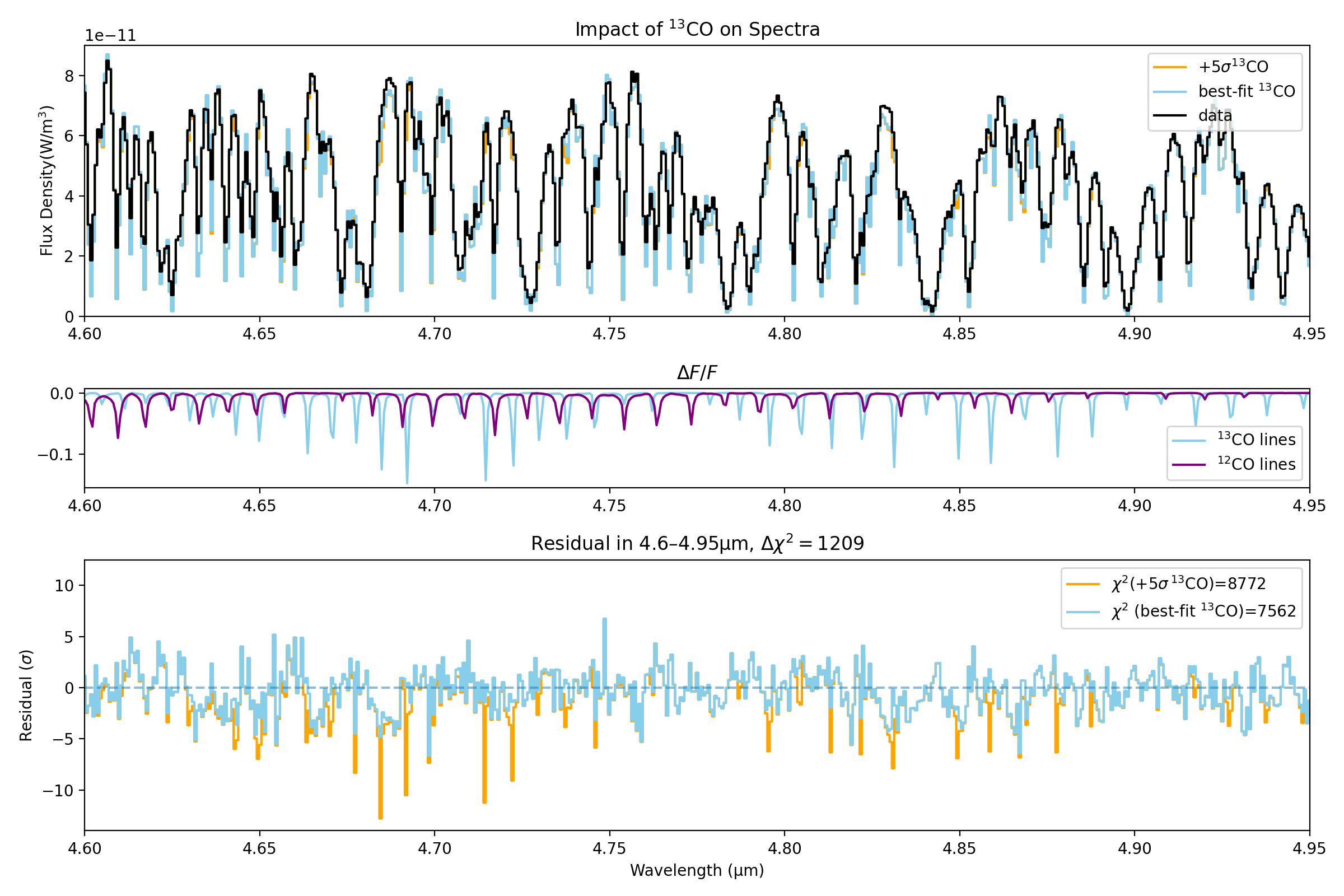}
    \caption{Top panel: The retrieved spectrum with the best-fit and the 5$\sigma$ upper limit $^{13}CO$ abundance; Middle panel: the change of the spectra after removing $^{13}$CO and 5\% $^{12}$CO opacity;  Bottom panel: the spectral difference between the two model spectra relative to the observational data uncertainty.}
    \label{fig:13cofig}
\end{figure*}

The CHIMERA retrieval gives a $\log[\frac{^{13}CO/^{12}CO} {(^{13}CO/^{12}CO)_{\textrm solar}}]$ of $-1.61^{+0.29}_{-0.39}$, which corresponds to log($^{13}$CO) of -8.58. 
However, the marginalized posterior distribution of log($^{13}$CO/$^{12}$CO) does not manifest a clear lower boundary.
To further study the impact of $^{13}$CO opacity on the modeling results, we use PICASO to model the spectra with the best-fit $^{13}$CO abundances (log($^{13}$CO)=-8.58) and with 5-$\sigma$ $^{13}$CO abundances (log($^{13}$CO)=-7.61) in  Figure \ref{fig:13cofig}.
In Figure \ref{fig:13cofig}, we plot the modeled spectra with two different $^{13}$CO abundances.
The residuals of the CHIMERA retrieval are within $7\,\sigma$ in the 4.6-4.96\,µm region. 
Upon the manual inspection of the elevated residuals, we conclude that we can rule out a $^{13}$CO enriched atmosphere with a log($^{13}$CO) of -7.61 because the additional $^{13}$CO opacity increases the maximum residuals to $10\,\sigma$ and increases the chi-squares by $\sim$1200.
In conclusion, we do not detect significant $^{13}$CO opacity in our data and place an upper limit of log($^{13}$CO) of less than -7.96, or $^{12}$CO/$^{13}$CO higher than 40.

\subsection{Search for possible trace species}\label{sec:trace}
We observe that there are numerous residuals in the model fit that exceed 5 $\sigma$ levels for both CHIMERA and Elf-Owl/PICASO models (e.g., Figure \ref{fig:retrievalspec}).
In several instances, the residuals exhibit wavelength-correlated behavior pointing to potential trace species that were not included in the model fit.
We, therefore, examined the detected molecules in Jupiter and Saturn's atmosphere, which are about 200-300\,K colder than \target, for potential molecules present in the atmosphere of \target.
The 3-5\,µm infrared spectra of Jupiter and Saturn show absorption features of a variety of trace molecular species, including \ce{PH3}(4.6-4.8\,µm)\citep{bezard1989}, \ce{GeH4}(4.737\,µm)\citep{bezard1989,giles2017}, \ce{AsH4}(4.704\,µm,4.728\,µm)\citep{bezard1989,noll1989,giles2017}, , and \ce{C2H2}($\sim$ 3.7\,µm, $\sim$ 4.6\,µm) \citep{ridgway1974,hitran2004}.
However, we do not detect strong deviations at the line positions of those possible molecular species based on the residuals of the best-fit models.
We find that the Elf-Owl model fitting residuals (data$-$model) have a  $\sim 20\sigma$ positive  deviation, which is higher than the averaged residuals (see Figure \ref{fig:retrievalspec}) at 4.55\,µm. 
We attribute the positive residual to the inclusion of \ce{CH3D} in the opacities used in PICASO. Specifically, the HAPI code \citep{hapi} was used to compute the opacity of \ce{CH4}, which automatically weights the isotopologues by their Earth abundances. For \ce{CH3D}, this is 1\% the abundance of \ce{CH4}\footnote{\url{https://hitran.org/lbl/2?6=on}}. On the other hand, although CHIMERA uses identical \ce{CH4} line lists as PICASO \citep{Hargreaves2020ApJS}, it only includes the main isotopologue. 
 \ce{CH3D} is known to have a distinct absorption feature at 4.55\,µm \citep[e.g.,][]{morley2018}. 
Therefore, the strong ($\sim$20 $\sigma$) positive residual of the Elf-Owl best-fit model at 4.55\,µm suggests that \target's atmosphere is likely to have less than 1\% \ce{CH3D/CH4}.
This result is consistent with CHIMERA retrieval's residuals that do not exhibit high residuals at 4.55\,µm because CHIMERA does not include \ce{CH3D} in the \ce{CH4} opacity calculation.

\section{Discussion}

\subsection{Importance of detected molecular abundances for atmospheric physics and chemistry of cool giant planets}

The presented 3-5\,µm spectrum of \target\ with JWST has demonstrated the value of R$\sim$3000 mid-infrared spectroscopy for atmospheric characterization of brown dwarfs.
In the cool Y-dwarf atmosphere, the major carbon-bearing species is \ce{CH4}. 
The 3-5\,µm spectrum simultaneously constrains \ce{CH4, CO2, and CO} which is useful for understanding the non-equilibrium chemistry in the Y-dwarf atmosphere.
 The best-fit Elf-Owl models suggest a moderate vertical mixing strength with a log(\kzz ($\textrm cm^2s^{-1}$)) of 4.
 Our fitted vertical mixing is comparable with the fitted vertical mixing strengths in the legacy Spitzer study of L and T dwarfs \citep{stephens2009}. More atmospheric studies of Y dwarfs will be necessary to probe the variations of vertical mixing strengths and the corresponding non-equilibrium chemical composition in these cool, methane-dominated atmospheres.

The acceptable fit of the non-equilibrium model fitting results by the Elf-Owl models also validates the \citet{mukherjee2022}'s non-equilibrium chemical model in the cool Y-dwarf atmospheric regime. 
The lack of \ce{PH3} absorption features in the 4.2-4.3\,µm region, where \ce{PH3} is predicted by non-equilibrium chemistry \citep{visscher2006}, suggests that our understanding of phosphorus chemistry is incomplete.

\subsection{Comparison to evolutionary models and potential binarity}
\begin{figure}
    \centering
\includegraphics[width=0.5\textwidth]{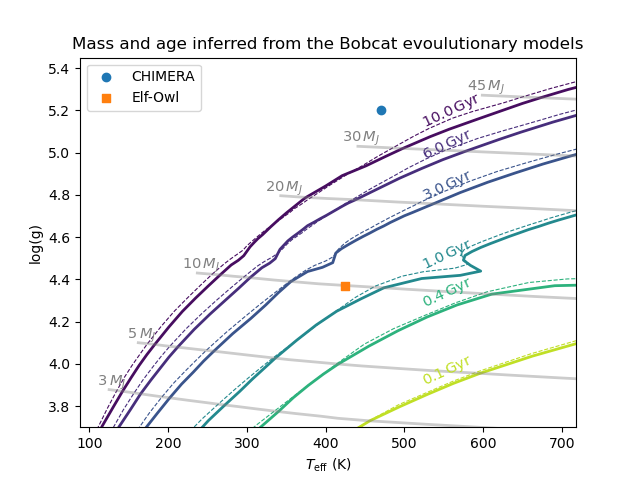}
    \caption{Based on the Sonora Bobcat evolutionary models, the fitted CHIMERA temperature and gravity implies an age of older than 10Gyr and mass of around 33 \mjup, or the fitted gravity is likely unphysically high (see \citealt{saumon2008}).
     The fitted Elf-Owl temperature and gravity indicates an age of 1.4\,Gyr and 10 \mjup. The colored contour lines show the age contour of a brown dwarf while the grey contour lines show the mass at an effective temperature and gravity. The solid and dashed lines represent the evolutionary model predictions for half-solar and solar metallicity objects respectively.}
    \label{fig:evolution}
\end{figure}

Brown dwarf evolutionary models \citep[e.g.,][]{phillips2020,marley2021} simulate the evolution of brown dwarf atmospheres with different metallicities and masses, and predict their radius and temperature changes with age.
The Sonora Bobcat evolutionary models \citep{marley2021} suggest that \target\ has a mass of 9.982 $\textrm M_{Jup}$ and an age of 1.4Gyr based on the fitted Elf-Owl temperatures and gravity.
The CHIMERA results suggest that the best-fit radius and gravity are 1.23 $\pm 0.01 \,\mathrm R_{Jup}$ and log(g) of 5.21$\pm 0.01$ respectively.
The fitted gravity by CHIMERA is higher than the expected gravity at 10\, Gyr by the evolutionary models (see Figure \ref{fig:evolution}).
Therefore, the fitted Elf-Owl grid-model solution, which has a lower metallicity than the CHIMERA retrieval, is more consistent with the evolutionary model prediction.
\added{We defer to Section \ref{sec:caveats} for a more detailed discussion on the impact of metallicity on the results.}
The inconsistent gravity between atmospheric retrieval frameworks and evolutionary models is not uncommon, and has been reported in the atmospheric study of other brown dwarfs \citep[e.g.,][]{zalesky2019,kitzmann2020}.

Based on the Bobcat brown dwarf evolutionary models, a brown dwarf with a $T_{\textrm eff}$ of 533\,K, solar or 2x solar metallicity, and a log(g) of around 5.08 has a radius of around 0.87 $R_{\textrm Jup}$ and $33M_{\textrm Jup}$.
The fitted radius is therefore about 40\% higher than the evolutionary model prediction.
One possible explanation for the discrepancy between our results and the evolutionary model radius is that \target\ is an unresolved equal-mass binary.
In the binary scenario, the radius of \target's single component will be equal to the square root of the best-fit radius, or 0.87 $R_{\textrm Jup}$.
A radius of 0.87 $R_{\textrm Jup}$ is then roughly  consistent with the evolutionary model radius at 10Gyr.

We also inspect if the fitted radial velocity is consistent with a potential binary scenario.
\added{We estimate the semi-major axis based on the following assumptions as to the inclination, mass, and semi-amplitude of the radial velocity of the potential binary.
We first assume that the possible inclinations of the binary system follows an isotropic distribution.
We assume the mass of each binary component is $33 M_{\textrm Jup}$ as indicated by the CHIMERA best-fit \teff\, and gravity.
Finally, we assume that the binary system has a semi-amplitude of $32.0 \pm 0.2$ km/s, which is the same as the CHIMERA best-fit radial velocity.
Based on these assumptions, our Monte Carlo estimation results suggest that such a potential binary system would have a semi-major axis of $0.0098^{+0.002}_{-0.006}$ \,au, or $20^{+4}_{-12} R_{\textrm Jup}$.}
This derived semi-major axis is smaller than the 0.5\,au upper limit estimated by \citet{defurio2023}. 
However, this order-of-magnitude estimation does not account for the radial velocity of the potential binary system and the uncertainty of the absolute wavelength calibration, which is below 14km/s \footnote{\url{https://jwst-docs.stsci.edu/jwst-data-calibration-considerations/jwst-data-absolute-wavelength-calibration}}. Future multi-epoch observations will be essential to detect potential radial velocity variations of \target.

\subsection{Caveats of atmospheric retrieval results}\label{sec:caveats}
Our CHIMERA retrieval results provide a decent fit to the 3-5\,µm spectrum with a reduced chi-squared of 5.4.
The retrieved gravity and radius are higher than the evolutionary model prediction. Here, we discuss the impact of potential bias in the retrieved gravity and radius on the results.

Our retrieved molecular abundances are positively correlated with gravity (see the figure in Appendix \ref{sec:appendix}). 
The correlation between gravity and abundance is because the photospheric pressure (P($\tau_\lambda$)=1) is proportional to the gravity and opacity ratio under the assumption of hydrostatic equilibrium:
\begin{equation}
    P(\tau = 1) \propto \frac{g}{opacity}
\end{equation}
Therefore, the photospheric pressure remains the same when both the volume mixing ratio and gravity increase by the same fractional amount.
Indeed, the molecular abundances except \ce{CO2} (see Section \ref{sec:composition} for the discussion of \ce{CO2} abundance) of the Elf-Owl model are about 0.5-1.0 dex lower than that of CHIMERA, while the gravity of Elf-Owl models is also about 0.8 dex lower than the CHIMERA retrieval (see Table \ref{tab:moleculelist}), similar to the expected correlation based on the analytical equation above.

 \begin{figure}
    \centering
    \includegraphics[width=0.5\textwidth]{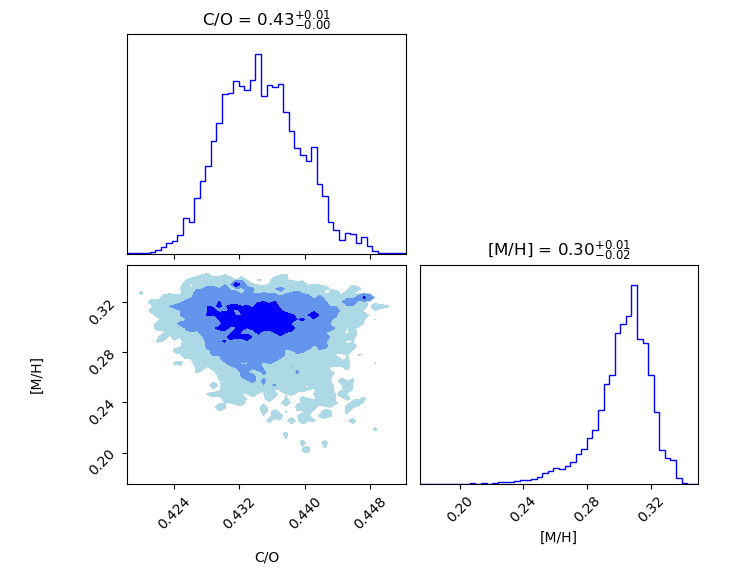}
    \caption{The metallicity and C/O ratios posterior distributions do not strongly correlate to each other.}
    \label{fig:metallicity}
\end{figure}

The interpretation of the effective temperature, metallicity, and gravity is less certain because of the degeneracy between gravity and molecular abundance. 
The best-fit Elf-Owl results present a low gravity, low temperature, and sub-solar metallicity scenario while the CHIMERA retrieval points to a high gravity, high temperature, and above-solar metallicity scenario.
The \ce{H2-H2} and \ce{H2-He} collision-induced absorption (CIA) \citep[e.g.,][]{saumon2012}, whose absorption features fall outside of the data wavelength coverage in this study, are sensitive to the photospheric pressure and gravity.
Therefore, further spectral analysis that extends to a wider wavelength coverage may help better constrain the gravity and distinguish the two scenarios.

The C/O ratio is less correlated to the fitted gravity because the abundance of carbon- and oxygen-bearing species are both positively correlated with gravity, therefore the C/O ratios should remain roughly the same even if the gravity changes. 
Indeed, the derived metallicity and C/O ratios from the MCMC samples do not strongly correlate with each other, as shown in Figure \ref{fig:metallicity}.
For the same reason, the relative abundances between two molecules in our retrieval results are also less correlated to the retrieved gravity.

\subsection{C/O ratio and metallicity in the solar neighborhood}\label{sec:cto}

 The C/O ratio and metallicity measured from planetary atmospheres are proposed as potential tracers of planet formation scenarios \citep{oberg2011,madhusudhan2011}.
 However,  connecting atmospheric abundances to the complex and under-constrained planet formation scenarios is challenging \citep[e.g.,][]{molliere2023}.
As a brown dwarf's mass range overlaps with the low-mass end of star formation and the high-mass end of planet formation, measurements of C/O and metallicity of a large sample of brown dwarfs will test if planet formation and star formation pathways could lead to different atmospheric properties.

We compare the derived C/O ratio and metallicity with the retrieval results in \citet{zalesky2019}.
In \citet{zalesky2019}, they use CHIMERA to retrieve the C/O ratios and metallicity of 14 late-T and Y dwarfs based on the HST spectra \citep{schneider2015}.
The retrieved C/O ratios range from 0.25 to 0.67 with a metallicity range from around -0.4 to +0.6 dex.
Therefore, our retrieved metallicity and C/O ratios of \target\ are consistent with that from \citet{zalesky2019}, as well as that of solar neighborhood stars of \citet{hinkel2014} (see Figure 5 in \citealt{zalesky2019}).
The retrieval results in \citet{zalesky2019} have C/O uncertainties of around 0.1 or around 20\% based on the HST 1.1-1.7\,µm spectra, and our results with the same retrieval framework has a C/O ratio uncertainty of 0.01 or 2\% level, which is an order-of-magnitude lower than the previous results. 
Our results illustrate that we can constrain the C/O ratios and metallicity to percent-level precision with the JWST spectroscopy.

\subsection{Implication of  \coisoratio\, ratio}
Recent {\coiso} detections in young giant planets and brown dwarfs \citep{zhang2021a,zhang2021b,gandhi2023} present an exciting opportunity to study the potential value of isotopologues in connecting the atmospheric properties to planet and star formation pathways.
A brown dwarf formed through gravitational collapse formation pathways should share similar isotopologues with the ISM values, which are set through processes including ion-exchange reactions \citep{langer1984},  isotope-selective photodissociation \citep{bally1982,vandishoeck1988, visser2009}, and gas-to-ice isotopologue partitioning \citep{Acharyya2007,smith2015}.
If a giant planet is formed in a disk, then additional processes, such as the dynamical CO depletion that varies across disk radius \citep{zhang2019}, could also be important and drive the isotopologue ratios  away from the ISM values \citep{yoshida2022}. 
Furthermore, the isotopologue ratios of a planet may also change with atmospheric evolution \citep[for reviews, see][]{nomura2022}.

Our results place a lower limit of the \coisoratio\, of 40.
\added{The isotopologue ratio lower limit of 40 is about 1$\sigma$ higher than that of the young giant planet  TYC 8998-760-1 b ($31^{+17}_{-10}$ \citealt{zhang2021a}).
The lower limit is consistent with the isotopologue ratio of the Sun ($86.8 \pm 3.9$ \citealt{scott2006,asplund2009}),  the brown dwarf 2MASS J03552337+1133437 ($97^{+25}_{-18}$ \citealt{zhang2021b}), and with the local ISM value within 1kpc of around 30-180 \citep[see Figure 5 in ][]{smith2015}.}
More modeling and observational studies of nearby stars and planets \citep[e.g.,][]{crossfield2019,zhang2021a,zhang2021b} are required to identify the dominating processes that shape the isotopologue ratio. 
Our JWST results demonstrate the potential of measuring the isotopologue ratio of cold giant planets for identifying potential trends and building up the sample size of planetary-mass objects with detected isotopologue.

\subsection{Comparison to \citealt{barrado2023} results} \label{sec:b23}
\citet{barrado2023} (hereafter B23) presented the JWST Mid-Infrared Instrument (MIRI) Medium-Resolution Spectrometer (MRS) 5-20\,µm spectra of \target\ and reported the detection of $^{15}\ce{NH3}$.
Therefore, both $^{15}$\ce{NH3} and $^{13}${CO} isotopologue constraints are available for the \target\ atmosphere.
B23 also reported the atmospheric properties derived from five different atmospheric retrieval frameworks.. Four of the five retrievals include both the MIRI spectra and the Hubble Space Telescope 1.1-1.7\,µm spectra.
We summarize the key results in B23 and compare them with our results in Table \ref{tab:miri}.
The derived gravity, metallicity, molecular abundances in B23 are consistent with our results within two-sigma levels.
B23's retrieval results prefer a lower effective temperature and a larger radius compared to our results.
The implied bolometric luminosity of B23, which is proportional to temperature $T^4$ and radius $r^2$, is similar to that of the Elf-Owl model within around 10\% level.
The temperature-pressure profile of B23 is similar to our results in the 1-10 bar region in Figure \ref{fig:contributionfunction}.
Therefore, the B23 results show similar chemical composition to both CHIMERA and Elf-Owl model results. The best-fit effective temperature of B23 is lower than that of both CHIMERA and Elf-Owl models, but the bolometric luminosity is similar with that of the Elf-Owl models.

\begin{table}[]
\centering
\caption{\centering Comparison of our modeling results to \citet{barrado2023}}\label{tab:miri}
\vspace{-0.4cm}
\begin{tabular}{cccc}
\hline \hline
 & \multicolumn{2}{|c|}{NIRSpec}  & MIRI+HST \\ 
Parameters                         & \multicolumn{1}{|c}{CHIMERA}  & \multicolumn{1}{c|}{Elf-Owl} & Retrieval \\ \hline
$T_{\textrm eff}$ (K) & $534^{+8}_{-25}$ & $425^{+0.46}_{-0.33}$ & $378^{+13}_{-18}$ \\
Radius  ($R_{\textrm Jup}$)                                     &  1.23    & $1.030 \pm 0.001$             & $1.37^{+0.26}_{-0.13}$                                         \\
    $\log(\textrm{g})$                                       &         $5.20 ^{+0.01}_{-0.02}$                   &     $4.38 \pm 0.01$ & $4.34^{+0.42}_{-0.88}$                           \\
$\left[M/H\right]$ & $+0.30 \pm 0.01$ &$-0.57 \pm 0.01$ & $-0.05^{+0.15}_{-0.27}$ \\
C/O & $0.46\pm0.01$ & $0.43\pm0.01$  & $0.21^{+0.45}_{-0.03}$\\
$\log(\ce{CH4}/\ce{H2O})$ & $-0.36 \pm 0.02$     & $-0.24$    & $-0.62^{+0.28}_{-0.30}$  \\
$\log(\ce{NH3}/\ce{H2O})$ & $-1.50^{+0.03}_{-0.02}$    & $-1.15$ & $-1.76^{+0.26}_{-0.30}$ \\ \hline 
\end{tabular}
\end{table}

\subsection{Comparison to \citet{leggett2021,cushing2021} studies} 

In this section, we will discuss our findings within the context of prior modeling studies concerning \target\ as reported by \citet{cushing2021} and \citet{leggett2021}.

In their study, \citet{leggett2021} assumed that \target\ is a binary system. They employed modified T-P profile models developed by \citep{tremblin2015,tremblin2019} to fit the Spectral Energy Distribution (SED) of \target. This SED encompassed data from the 1.1-1.7\,µm Hubble Space Telescope spectrum, the Spitzer [3.6] and [4.5] photometric bands, as well as the WISE W1 and W2 photometric bands. Their analysis yielded a \teff\, of 375\,K, log(g) of 4.0, [M/H] of -0.5, and log(\kzz) of 7.
In contrast, \citet{cushing2021} explored both cloudless and cloudy models to fit the SED of \target. For a single-object scenario, their best-fit cloudless model suggested a half-solar metallicity, 0.25 times the solar C/O ratio, a \teff\, of 350\,K, and a log(g) of 4.0. 
The best-fit cloudy model, on the other hand, exhibited a \teff\, of 275\,K and log(g) of 4.5. 
When considering the possibility of an unresolved binary system, their best-fit Sonora Bobcat binary model indicated a \teff\,  of 300\,K and 350\,K, log(g) of 4.0, metallicity of -0.5 and -0.5 dex, and C/O ratios of 0.5 and 0.25 times the solar value, respectively.

One significant factor contributing to the disparities between our results and those of previous studies likely arises from differences in the wavelength coverage of the dataset. Previous modeling efforts primarily centered on fitting models to the near-infrared spectrum and mid-infrared photometry. In contrast, our analysis focused on fitting the models to the 3-5\,µm spectrum. Given that the 3-5\,µm spectrum is particularly sensitive to the abundance of \ce{CO2}, \ce{CH4}, \ce{CO}, and \ce{H2O}, our findings are expected to provide a more precise estimate of the C/O ratio compared to prior studies.

It's important to note that the near-infrared spectrum probes pressure levels higher than those probed by the mid-infrared spectrum, as illustrated in Figure \ref{fig:contributionfunction}. Consequently, the T-P profile derived from the previous modeling studies should offer a more accurate estimate of the temperature at pressures greater than around twenty bars.
A combined analysis incorporating archival near-infrared and mid-infrared spectra would provide a more comprehensive and less biased view of the  temperature-pressure profile of \target.

\section{Conclusions}
Our study marks one of the first in-depth atmospheric investigations of a Y-dwarf using JWST NIRSpec spectroscopy. 
Our rich results provide a glimpse of the powerful capabilities of JWST in revolutionizing our understanding of cold giant planet atmospheres.
We summarize our key findings as follows:

\begin{enumerate}
    \item We present the JWST NIRSpec/G395H 2.880-5.142\,µm spectrum of WISE 1828. Our moderate spectral resolution (R$\sim$ 2700) spectra of \target\ cover $\sim$47 \% of the bolometric luminosity of Y-dwarf \target\ with a peak signal-to-noise ratio exceeding 90.

    \item We utilize two complementary atmospheric modeling tools, the CHIMERA atmospheric retrieval framework and the Elf-Owl radiative-convective equilibrium grid models, to characterize the atmospheric structure and composition of WISE 1828. 
    
    \item We find that the best-fit Elf-Owl grid model has an effective temperature of $425^{+0.46}_{-0.33}$\,K, log(g) of $4.38 \pm 0.01$, log(\kzz) of $4.65 \pm 0.04$, C/O ratio of 0.46 $\pm 0.01$, and metallicity [M/H] of $-0.57 \pm 0.01$. 
    The CHIMERA retrieval has an effective temperature of $534^{+8}_{-25} $K, metallicity of $+0.30^{+0.01}_{-0.02} $ dex, and a C/O ratio of $0.43 \pm 0.01$.
    The reduced chi-squared values of the best-fit CHIMERA and Elf-Owl models are 5.4 and 13.2 respectively.
    While the CHIMERA retrieval has a lower reduced chi-squared value, the fitted gravity of the Elf-Owl model is more consistent with the Bobcat evolutionary model prediction.
    
    \item Based on the CHIMERA retrieval results, we reported the detection and bounded abundance constraints of \ce{CH4, H2O, CO, NH3, H2S, and CO2} molecules. 
    We report the first measurement of the \ce{H2S} abundance in a Y-dwarf atmosphere. Our residual analysis finds no evidence of trace molecular species \ce{PH3, GeH4, C2H2, AsH4}.We do not detect {\coiso} in the G395H data with CHIMERA. We place a lower limit in the $^{12}$CO/$^{13}$CO isotopologue ratios of 40. We derive the elemental abundances of [C/H], [N/H], [O/H], and [S/H]. We find qualitative evidence of a lower-than-nominal \ce{CH3D} abundance (i.e., \ce{CH3D/CH4} $<$ 1\%) in \target\ based on the Elf-Owl best-fit model's residuals at 4.55\,µm.

    \item The CHIMERA retrieval and Elf-Owl model fitting results demonstrate that the capability of JWST in characterizing Y-dwarf atmosphere to percent-level precision of C/O ratio and metallicity. We caution that the metallicity is highly model dependent while the fitted C/O derived from the CHIMERA retrieval and Elf-Owl models agree well to each other within 7\%.

    \item We attribute the difference in the forward modeling and the atmospheric retrieval results to the degeneracy between gravity and molecular opacity. We derived relative abundances of \target\ that are less dependent on the fitted gravity parameter in the atmospheric models. Future observations with a wider wavelength coverage will be useful to disentangle the degeneracy and provide a more accurate estimate of the effective temperature, elemental abundances, metallicity, and gravity of WISE 1828.
    
    \item The retrieved radius and gravity are both higher than that of the brown dwarf evolutionary models. One possible explanation is that \target\ is a tightly bound binary pair that share similar temperatures. We discuss the implication of the potential binarity based on the fitted results and evolutionary models. Our crude estimation of the semi-major axis is consistent with previous studies.

\end{enumerate}

\section*{Acknowledgement}
B.L. would like to acknowledge the funding support from NASA through the JWST NIRCam project though contract number NAS5-02105 (M. Rieke, University of Arizona, PI). B.L. would like to thank David Barrado, Polychronis Patalpis, Ehsan Gharib-Nezhad, and Paul Mollière for the useful discussions.
T. R. and T.G. were supported via NASA WBS 411672.07.05.05.03.02.
A.Z.G. acknowledges support from Contract No. 80GSFC21R0032 with the National Aeronautics and Space Administration. N.E.B acknoledges support from NASA’S Interdisciplinary Consortia for Astrobiology Research (NNH19ZDA001N-ICAR) under award number 19-ICAR19\_2-0041.
The JWST data presented in this article were obtained from the Mikulski Archive for Space Telescopes (MAST) at the Space Telescope Science Institute. The specific observations analyzed can be accessed via \dataset[DOI: 10.17909/jfy8-dm19]{https://doi.org/10.17909/jfy8-dm19}.

Software: PICASO \citep{batalha2019,mukherjee2023}, Ultranest \citep{Ultranest}, CHIMERA \citep{line2015}, AstroPy \citep{astropy:2013,astropy:2018,astropy:2022}, Pyphot \citep{pyphot}

\begin{figure*}
    \includegraphics[width=1.0\textwidth]{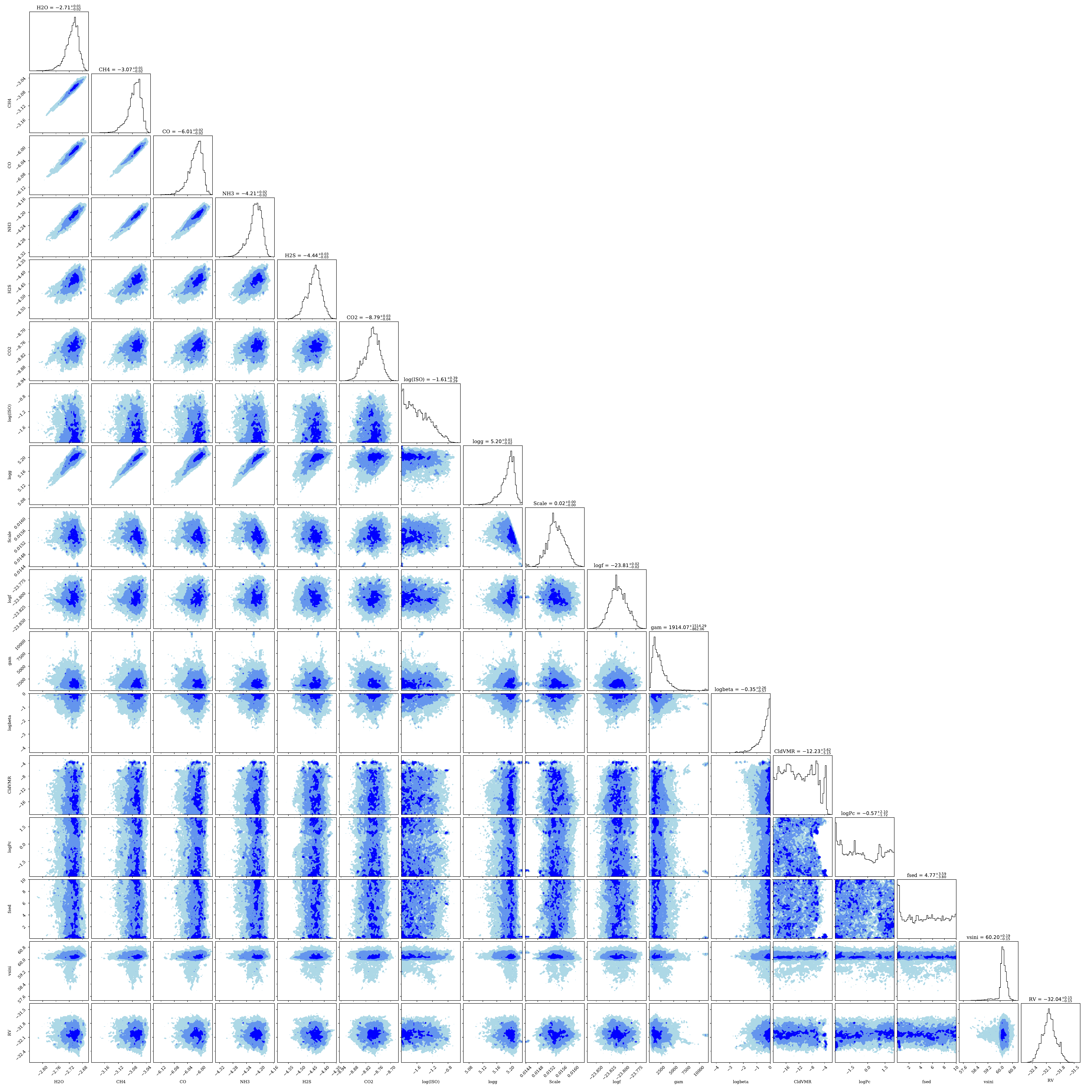}
    \caption{The corner plot of CHIMERA retrieval results. The parameters for TP profiles are excluded for clarity. See text for the explanation of the listed parameters.}\label{fig:chimera-corner}
\end{figure*}

\appendix \label{sec:appendix}
In Figure \ref{fig:chimera-corner}, we list out the full posterior distribution for the CHIMERA retrieval framework. The parameter \ce{H2O,CH4,CO, NH3,H2S,CO2,Na} are the logarithmic mixing ratios of molecules, log(ISO) is the logarithmic \ce{^{13}CO}/\ce{^{12}CO} isotopologue ratio relative to the solar value, $\log(g)$ is the logarithmic value of surface gravity in cgs unit, $ Scale$ is the squared value of the ratio of radius over distance in unit of ($R_{\textrm Jup}/pc)^2$), $\log(f)$ is the variance of inflated noise, $ gam$ is the smoothing hyperparameter for the TP profile \citep[i.e., Equation 5 in][]{line2015}, $\textrm logbeta$ is the parameter for the inverse gamma distribution for the hyperparameter $\gamma$, $\textrm CldVMR$ is the logarithmic volume mixing ratio at the cloud base, $\log(P_c)$ is the cloud base pressure, $f_{\textrm sed}$ is the cloud sedimentation efficiency \citep{ackerman2001}, $v\sin i$ is the rotational velocity in $\textrm kms^{-1}$, and $V_{\textrm rad}$ is the radial velocity in $\textrm kms^{-1}$.

\bibliographystyle{aasjournal}



\end{document}